\documentclass[12pt]{article}
\let\section=\subsection     \let\subsection=\subsubsection                
\usepackage{graphicx,amsmath,amssymb,bm}

\newcommand{\kf}{k_{\text{F}}}
\newcommand{\be}{\begin{equation}}
\newcommand{\ee}{\end{equation}}
\newcommand{\q}{{\bf q}}
\newcommand{\qp}{{\bf q'}}
\newcommand{\pp}{{\bf p}+{\bf p'}}
\newcommand{\ip}{{\bf p''}}
\newcommand{\la}{\Lambda}
\newcommand{\spincross}{{\bm \sigma}_1 \times {\bm \sigma}_2}

\begin{document}
\begin{center}
{\large \bf Renormalization Group Methods for the}\\[2mm]
{\large \bf Nuclear Many-Body Problem}\\[5mm]
A.~Schwenk$\,^{a}$, B.~Friman$\,^{b}$ and G.E.~Brown$\,^{c}$ \\[5mm]
{\small \it $^{a}$Department of Physics, The Ohio State University,
Columbus, OH 43210 \\
$^{b}$Gesellschaft f\"ur Schwerionenforschung,
Postfach 110552, 64220 Darmstadt \\
$^{c}$Department of Physics and Astronomy,
State University of New York, \\ Stony Brook, NY 11794-3800 \\[8mm] }
\end{center}

\begin{abstract} \noindent
The application of renormalization group (RG) methods to
microscopic nuclear many-body calculations is discussed. We present
the solution of the RG equations in the particle-hole channels for
neutron matter and the application to S-wave pairing. Furthermore,
we point out that the inclusion of tensor and spin-orbit forces leads 
to spin non-conserving effective interactions in nuclear matter.
\end{abstract}

The solution to the many-body problem for systems of strongly
interacting particles, such as finite nuclei and nuclear matter,
can often be facilitated, by making a judicious use of the
separation of low- and high-energy scales. One introduces a
truncated Hilbert space (model space), where the particles are
restricted to low energies. These are the so-called ``slow'' modes.
The operators and the degrees of freedom in the truncated space
must be renormalized to account for intermediate excitations to
states outside the model space, the ``fast'' modes. The operators
of interest include the effective interaction 
and various transition operators, e.g., the axial current. 
This procedure defines an effective theory, which is equivalent to 
the full theory at low energies. The RG method provides a 
systematic way to compute the effective operators of particles 
in the truncated space.

There are several advantages of working in a truncated space. By
integrating out the high-energy modes, the strong short-range
repulsion of realistic nucleon-nucleon forces is tamed, and the
resulting effective interaction is model independent, if it acts
only at energies that are constrained by the scattering
data~\cite{Vlowk1,Vlowk2,Vlowk3,Vlowk4}. Due to the separation 
of scales, it can usually be achieved that the effective operators 
in the model space are energy-independent. 
For finite nuclei, the model space concept has been used for many
years in the derivation of effective shell model interactions, where 
the truncated space contains the low-lying shells near the Fermi
energy.

In this first application of RG methods to effective nuclear
interactions, we consider infinite neutron matter. The model space 
consists of the single particle states near the Fermi
surface. In neutron matter, the effects of tensor as well as
three-body forces are reduced compared to symmetric nuclear matter,
since they do not act in the S-wave. Therefore, as a first
approximation we neglect these forces in this calculation. In
Section 4, we briefly discuss some novel consequences of tensor and
spin-orbit forces in a nuclear medium.
\begin{figure}[t]
\begin{picture}(125,75)
\put(0,0){\includegraphics[height=7.5cm,width=12.5cm,clip=]{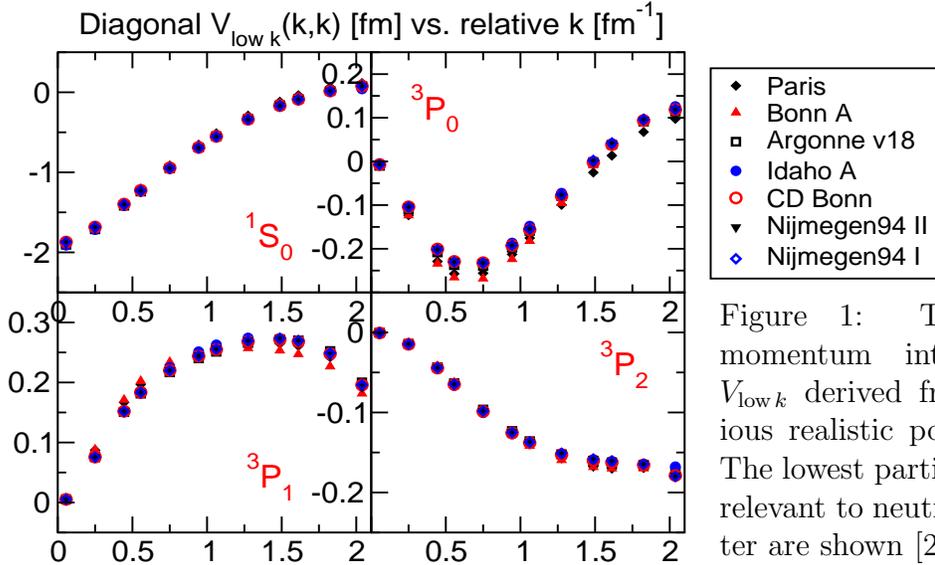}}
\put(95,18){\parbox{44mm}{\caption{The low momentum interaction
$V_{\text{low}\,k}$ derived from various realistic potentials. The
lowest partial waves relevant to neutron matter are
shown~\cite{Vlowk2,Vlowk3,Vlowk4}.\label{vlowkdiag}}}}
\end{picture}
\end{figure}

\section{Two-body low momentum interaction}

As input to the many-body calculation we take the two-body low
momentum potential $V_{\text{low}\,k}$~\cite{Vlowk1,Vlowk2,Vlowk3,Vlowk4},
which is derived from realistic nucleon-nucleon interaction models
by integrating out the high momentum modes in the sense of the RG.
Here, we briefly summarize the properties of $V_{\text{low}\,k}$
and refer to the original papers and Ref.~\cite{rg2002} for
details. The RG decimation to low momenta is constructed such that
$V_{\text{low}\,k}$ is energy-independent and reproduces the
half-on-shell $T$ matrix for momenta below a cutoff $\la$. For
$k',k < \la$, we thus have
\begin{equation}
T(k',k;k^{2}) = V_{\text{low}\,k}(k',k) +  \frac{2}{\pi} \;
\mathcal{P} \int_{0}^{\Lambda} \frac{V_{\text{low}\,k}(k',p) \;
T(p,k;k^{2})}{k^{2}-p^{2}} \; p^{2} dp .
\label{vlowkeq}
\end{equation}
Consequently, $V_{\text{low}\,k}$ is phase shift equivalent to the
original nucleon-nucleon interaction. For values of the cutoff, 
$\la \sim 2.1 \, \text{fm}^{-1}$ (the scale up to which the potential 
models are fitted to empirical phase shifts), one finds the 
same low momentum potential for the various realistic nucleon-nucleon 
interactions, i.e., the resulting potential is model independent, as 
demonstrated in Fig.~\ref{vlowkdiag}.

In $V_{\text{low}\,k}$, the model-dependent high momentum (short
distance) modes have been integrated out. This procedure tames the
strong short-range repulsion in the original interaction. Thus, the
low momentum interaction $V_{\text{low}\,k}$ can, in contrast to the
bare interaction, be used directly in many-body calculations.
Furthermore, since the states above the cutoff $\la$ are included in
$V_{\text{low}\,k}$ and those below are not, the cutoff acts much
like the Pauli-blocking operator in the Brueckner $G$ matrix~\cite{IIpaper}.
By analyzing the angle-averaged Pauli blocking operator, one finds
an effective cutoff slightly larger than the Fermi momentum. Here
we employ a density-dependent cutoff for $V_{\text{low}\,k}$,
$\la_{V_{\text{low}\,k}} = \sqrt{2}\,\kf$. Due to the separation of
short distance and long distance scales, $V_{\text{low}\,k}$ is
almost independent of the cutoff in the $T=1$ channel~\cite{Vlowk2}
for $0.7$ fm$^{-1} \lesssim \la \lesssim 2.1$
fm$^{-1}$. Thus, the exact value of
the cutoff is not crucial for pure neutron matter.

\section{Renormalization group approach to Fermi liquids}

\begin{figure}[t]
\begin{picture}(125,43)
\put(2,0){\includegraphics[scale=0.26,clip=]{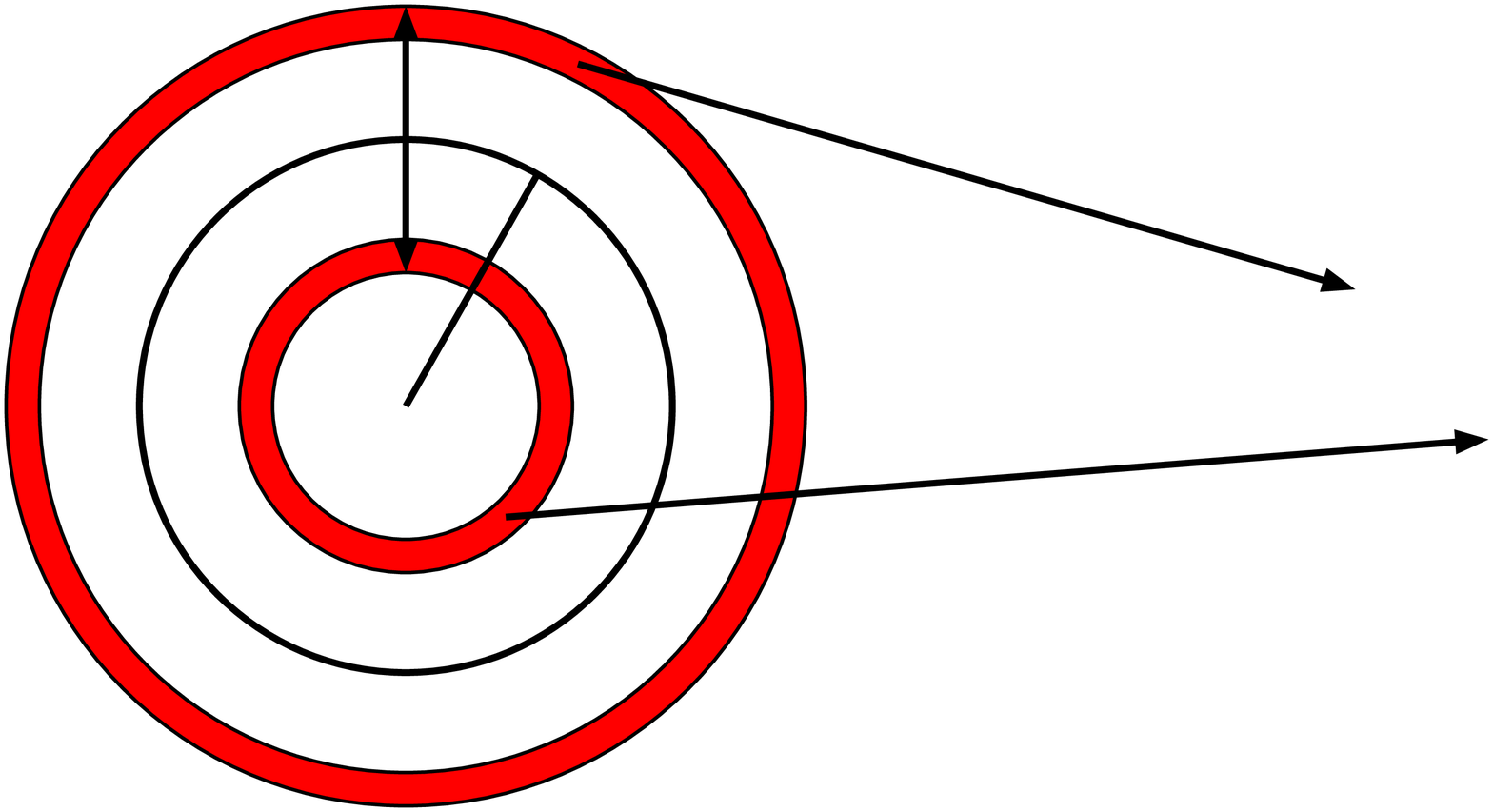}}
\put(28,26){$\kf$}
\put(18,29){$\la$}
\put(18,34){$\la$}
\put(46,12){fast hole}
\put(41,35){fast particle}
\put(57,0){\includegraphics[scale=0.72,clip=]{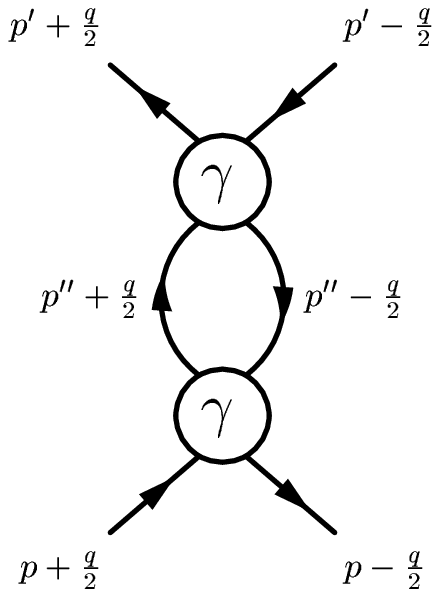}}
\put(92,6){\includegraphics[scale=0.72,clip=]{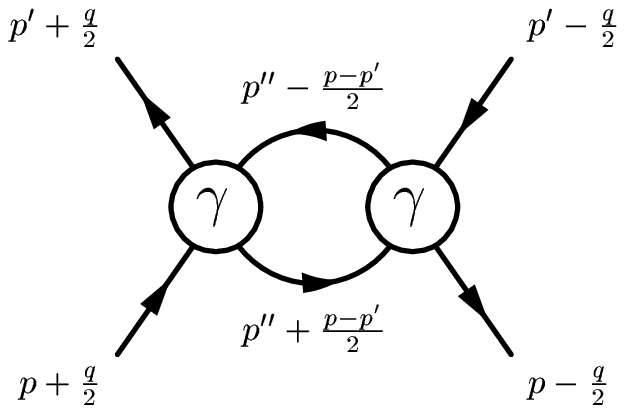}}
\end{picture}
\caption{The one-loop contributions to the RG equation, where $\gamma(\la)$
denotes the running four-point vertex. For particles on the Fermi
surface, the four-point vertex at zero energy transfer $\omega$
is given by $a(\la)=\gamma(\omega=0,\la)$. On the left, the momentum
shells which are integrated out at every step $\la$ are shown.\label{zsandzsp}}
\end{figure}

We now discuss the RG approach to the many-fermion problem. As
originally proposed by Shankar~\cite{Shankar}, one separates the
slow modes from the fast ones in a many-fermion system by
introducing a momentum cutoff relative to the Fermi momentum. At
zero temperature, the phase space for fast modes is then
characterized by the occupation factors
\begin{equation}
n_{\bf p}(\la) = \Theta(\kf - \la - |{\bf p}|) \quad \text{and} \quad 1-n_{\bf
p}(\la) = \Theta(|{\bf p}|- (\kf + \la)) .
\end{equation}
The slow modes that make up the model space are then located in the
complementary shell of thickness $2\la$ around the Fermi surface.
Starting at a large cutoff ($\la=\kf$), where the effective
interaction in the model space is assumed to be given by the
free-space low momentum interaction $V_{\text{low}\,k}$, it is
successively renormalized by including the contributions of fast
intermediate states lying in a thin shell between $\la -
\delta\la$ and $\la$. For $\delta\la\to 0$, one obtains a
differential (RG) equation for the renormalization of the effective
interaction as a function of the cutoff. As $\la$ is decreased,
more and more modes are shifted from the model space into the
effective interaction, in such a way that the low energy scattering
amplitude remains invariant. At the same time the single-particle
degrees of freedom are gradually converted into quasiparticles. As
$\la \to 0$, the effective interaction equals the scattering
amplitude for quasiparticles on the Fermi surface, since all modes
have been integrated out. To one-loop order, the RG
equation in the particle-hole channels, which play a special role
in Fermi liquid theory, is at zero temperature given by~\cite{IIpaper}
\begin{align}
\frac{d}{d \Lambda} a(\q,\qp;\la) &= z_{\kf}^2 \:
\frac{d}{d \Lambda} \biggl\{ \;
g \int\limits \frac{d^3 \ip}{(2 \pi)^3} \:
\frac{n_{\ip+\q/2}(\la)-n_{\ip-\q/2}(\la)}{\varepsilon_{\ip+\q/2}
-\varepsilon_{\ip-\q/2}} \biggr\} \nonumber \\[1mm]
&\times a\bigl(\q,\frac{\pp}{2}+\frac{\qp}{2}-\ip;\la\bigr) \;
a\bigl(\q,\ip-\frac{\pp}{2}+\frac{\qp}{2};\la\bigr) \nonumber \\
&- \text{exchange channel} \: \biggl\{ \q \leftrightarrow \qp
\biggr\} .
\label{rge}
\end{align}
The contributions from the direct and the exchange channel are
shown diagrammatically in Fig.~\ref{zsandzsp}. When only one
channel is considered, the one-loop RG equation is exact in the
sense that it is equivalent to the corresponding scattering
equation. Within the RG method, the scattering amplitude remains
antisymmetric at any-loop order when both particle-hole channels
are included.

We choose a scattering geometry, where the external particles are restricted
to the Fermi surface, $|{\bf p}^{(\prime)}\pm \q/2|= \kf$. Then, the
scattering amplitude depends only on two angles, or equivalently the magnitude
of the two momenta $|\q|$ and $|\qp| = |{\bf p}-{\bf p}^\prime|$. The direct
and exchange channels couple in the RG equation, Eq.~(\ref{rge}).
Microscopically the RG in the particle-hole channels includes particle-hole
bubbles and ladders as well as vertex corrections. At this point we include
only the high-momentum ladder diagrams summed in $V_{\text{low}\,k}$ in the
particle-particle channel. Finally, the effective interaction of particles on
the Fermi surface is obtained by setting $|{\q}| = 0$ in
Eq.~(\ref{rge})~\cite{RGnm}.  The resulting effective interaction includes the
so-called induced interaction.

\section{Neutron matter and S-wave pairing}

We solve the RG equation, Eq.~(\ref{rge}), for neutron matter (for
details see Ref.~\cite{RGnm}). The evolution of the effective mass
as well as an approximate treatment of the renormalization of the
quasiparticle strength is included in the flow. We note that the RG
approach does not require a truncation in Landau $l$ of the Fermi
liquid parameters. The resulting Landau parameters are given in
Ref.~\cite{RGnm}.

\begin{figure}[t]
\begin{picture}(125,65)
\put(0,0){\includegraphics[scale=0.34,clip=]{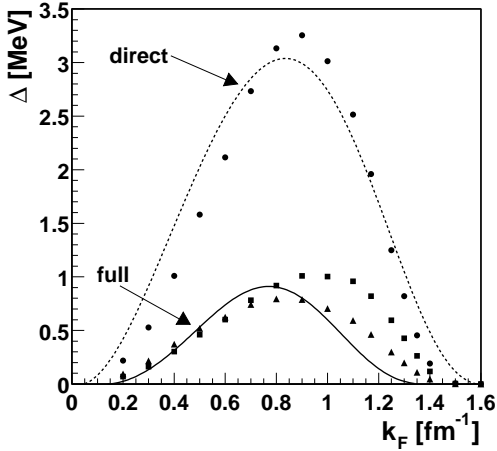}}
\put(72,31){\parbox{68mm}{\caption{The $^1$S$_0$ superfluid gap versus
the Fermi momentum $\kf$. The dots denote the gap obtained using
$V_{\text{low k}}$ only, whereas the squares and the triangles are
computed with the full RG solution using different approximations
for the quasiparticle strength~\cite{RGnm}. The dashed line is
obtained by solving the BCS equation with various bare
interactions~\cite{LomSch}, while the solid line includes
particle-hole polarization effects~\cite{WAP}.\label{1s0gaps} }}}
\end{picture}
\end{figure}
While the Fermi liquid parameters are defined in the forward
scattering limit, the RG procedure yields the quasiparticle
scattering amplitude also for finite scattering angles. The latter
can be used to compute the S-wave superfluid gap. In weak coupling 
BCS theory,
\begin{equation}
\Delta = 2 \, \varepsilon_{\text{F}} \exp\biggl( \frac{1}{\langle
\mathcal{A} \rangle} \biggr) ,
\end{equation}
where $\langle \mathcal{A} \rangle$ denotes the angle averaged
scattering amplitude on the Fermi surface, we find the pairing gap
shown in Fig.~\ref{1s0gaps}. The S-wave gap is strongly suppressed,
from $3.3\, \text{MeV}$ to $0.8 \, \text{MeV}$ at maximum, due to
particle-hole screening in the many-body medium. This agrees well
with the results obtained in the polarization potential model by
Wambach {\it et al.}~\cite{WAP}.

At higher densities, neutrons form a P-wave ($^3$P$_2$--$^3$F$_2$)
superfluid. For fermions interacting by means of a delta function
force, the second order particle-hole contributions to the P-wave 
scattering amplitude are attractive for back-to-back configurations 
on the Fermi surface, and thus one expects the P-wave gap to increase. For a
quantitative analysis of pairing in this channel, however, the effects 
of tensor and spin-orbit forces in the medium must be included.

\section{Tensor and spin-orbit forces in the nuclear medium}

Here we briefly discuss the noncentral parts of the effective
two-nucleon interaction. In a nuclear medium, new spin-dependent
interactions are induced by the presence of other particles nearby.
The polarization of the Fermi sea leads to contributions, which
depend on the two-body cm momentum ${\bf P}$ in the rest frame of
the many-body system. In the long-wavelength Landau limit, two of these
new operators survive, a modified spin-orbit term $(\spincross)\cdot
(\qp\times{\bf P})$, which does not conserve two-particle spin,
and a tensor term $S_{12}({\bf P})$. Furthermore, relativistic
corrections related to the transformation from the two-body cm
frame to the rest frame of the many-body system also contribute to
such nonstandard operators~\cite{rel}. Neither of these effects
are included in conventional many-body calculations, where the two-body
interaction is treated in the independent pair approximation. A
calculation of the effective tensor and spin-orbit interactions in
nuclear matter is in progress~\cite{tensor}.

\vspace*{-1mm}
\subsection*{Acknowledgements}
We thank S. Bogner, H. Feldmeier, R. Furnstahl, T. Kuo and
D.-O. Riska for stimulating discussions. AS is supported
by the Ohio State University through a Postdoctoral
Fellowship and the NSF under Grant No. PHY-0098645. The work of GEB is
supported by the US DOE Grant No. DE-FG02-88ER40388.

\end{document}